\documentclass[twocolumn,showpacs,preprintnumbers,amsmath,amssymb, superscriptaddress]{revtex4}

\usepackage{graphicx}
\usepackage{dcolumn}
\usepackage{bm}

\usepackage{amsmath,amstext,amssymb}
\usepackage{amsmath}

\begin{document}

\title{Coulomb effects in semiconductor quantum dots}

\author{Norman Baer}
\affiliation{Institute for Theoretical Physics, University Bremen,
             P.O. Box 330 440, 28334 Bremen, Germany}
\author{Paul Gartner}
\affiliation{Institute for Theoretical Physics, University Bremen,
             P.O. Box 330 440, 28334 Bremen, Germany}
\affiliation{National Institute of Materials Physics, PO Box MG-7,
			 Bucharest-Magurele, Romania}			 
\author{Frank Jahnke}
\affiliation{Institute for Theoretical Physics, University Bremen,
             P.O. Box 330 440, 28334 Bremen, Germany}

\date{\today}
 
\begin{abstract}
Coulomb correlations in the optical spectra of semiconductor quantum
dots are investigated using a full-diagonalization approach. The
resulting multi-exciton spectra are discussed in terms of the
symmetry of the involved states. Characteristic features of the
spectra like the nearly equidistantly spaced $s$-shell emission lines
and the approximately constant $p$-shell transition energies are
explained using simplified Hamiltonians that are derived taking into
account the relative importance of various interaction contributions.
Comparisons with previous results in the literature and their 
interpretation are made.  
\end{abstract} 

\maketitle

\section{Introduction}
\label{Introduction}

Recent experimental investigations of optical spectra from individual
self-assembled semiconductor quantum dots   (QDs) \cite{Bayer:02}
renewed the interest in fundamental quantum-mechanical effects of
confined interacting few-carrier systems and their theoretical
description
\cite{Barenco:95,Hawrylak:99,Dekel:00a, Hohenester:00,Brasken:01}. The
few-carrier systems in individual QDs are interesting candidates for
potential applications in quantum optics like sources of single
photons or entangled photon pairs \cite{Moreau:01}.

Apart from the experimental relevance, the problem of few charge
carriers in the discrete states of a given confinement potential is
also a paradigm for an interacting few-carrier system, that can be
solved without further approximations. A discussion of the exact
(but complicated) results in terms of simplified pictures is
therefore desirable.

In the past, electronic states and the resulting dipole transition have
been calculated in box-like confinement potentials using exact
diagonalization \cite{Barenco:95} as well as in a two-dimensional
system with harmonic confinement potential \cite{Hawrylak:99} with a
limited number of configurations considered in the diagonalization.
Furhermore, configuration interaction calculations for a numerically
determined strain-induced confinement potential \cite{Brasken:01} and
density functional calculations of the energy level structure and
luminescence spectra \cite{Nair:00} have been performed.

In this paper we study the influence of Coulomb correlations on the
optical spectra of QDs. Starting from the localized single-particle
states in flat, cylindrically symmetric QDs, described in the envelope
function picture, the interacting states are obtained from the full
diagonalization of the Hamiltonian including Coulomb interaction. From these
states, optical emission spectra are computed for an increasing number of
excitons assumed to be in their (interacting) ground states. Even though the
Hamiltonian considered here is the same as in Ref.\cite{Hawrylak:99}, the full
diagonalization results differ in several respects from the approximate
picture obtained there.

Despite the complexity of the spectra, characteristic features appear that can
be explained in terms of simplified Hamiltonians. These are obtained from the
original Hamiltonian by retaining dominant terms and neglecting less
important ones, such that, on the one hand, the essential spectral
features are preserved and, on the other hand, analytic results can be
deduced.  The aim is to get a more intuitive picture of the influence of
Coulomb effects and to provide an alternative to a numerically demanding full
diagonalization approach. The trade-off between accuracy and simplicity can be
reached in several ways and we give here two examples whose merits and
shortcomings are assessed against the full calculation result.

A first case is represented by a diagonal Hamiltonian, which is obtained
by retaining only the direct and exchange terms, and which describes fairly
well the main spectral features. The role of the Coulomb exchange interaction
in the ladder-like structure of the lower part ($s$-lines) of the emission
spectrum, recently observed in Refs.~\cite{Dekel:98, Dekel:00a} can be easily
accounted for in this description.

For explaining the absence of a similar ladder structure in the upper part
($p$-lines) of the spectrum a more careful handling of the Coulomb interaction
is required. This is done in the second simplified approach described by
an adiabatic Hamiltonian. One encounters here the so-called 'hidden
symmetry'. This property was discussed in a series of papers
\cite{Wojs:96, Hawrylak:96, Bayer:00, Hawrylak:03}. Nevertheless,
the proof of the argument makes use of the assumption that one deals with a
single degenerate shell, while in fact several, not weakly interacting shells
are always present. Therefore it is by no means clear if the 'hidden symmetry'
result holds, and if it does why and in what form. We show that in the
adiabatic approach, and due to certain peculiarities of the problem, the
'hidden symmetry' can be recovered, albeit with the parameters  renormalized
by the interaction between the shells.
The discussion of the conditions in which this happens sheds light
on the limits of validity of the 'hidden symmetry' argument.


\section{Hamiltonian}

To describe the system of QD electrons and holes under the influence
of Coulomb interaction we use the Hamiltonian $H = H_0+ H_{Coul}$ in
the envelope-function approximation,
\begin{align}\label{eq:ehHamiltonian}
 H_0=&\sum_{i\sigma} \varepsilon^e_i e^{\dag}_{i \sigma}e_{i \sigma} +
      \sum_{i\sigma} \varepsilon^h_i h^{\dag}_{i \sigma}h_{i \sigma} \quad,
      \nonumber \\
 H_{Coul}=
      & \frac{1}{2} \sum_{ijkl \atop \sigma \sigma'} V^{ee}_{ij,kl} \;
      e^{\dag}_{i \sigma}  e^{\dag}_{j \sigma'} e_{k \sigma'} e_{l \sigma}
\nonumber \\
    + & \frac{1}{2} \sum_{ijkl \atop \sigma \sigma'} V^{hh}_{ij,kl} \;
      h^{\dag}_{i \sigma}  h^{\dag}_{j \sigma'} h_{k \sigma'} h_{l \sigma}
\nonumber \\
    - &             \sum_{ijkl \atop \sigma \sigma'} V^{he}_{ij,kl} \;
      h^{\dag}_{i \sigma}  e^{\dag}_{j \sigma'} e_{k \sigma'} h_{l \sigma}
\quad.
\end{align}
Here $e_{i \sigma}$ ($e^{\dag}_{i \sigma}$) are annihilation
(creation) operators of electrons with spin $\sigma$ in the
one-particle orbital states $|i\rangle$ of energy
$\varepsilon^e_i$. The corresponding operators and single-particle
energies for holes are $h_{i \sigma}$ ($ h^{\dag}_{i \sigma}$) and
$\varepsilon^h_i$, respectively.  The Coulomb interaction matrix
elements are given by
\begin{multline}\label{eq:Vijkl}
 V^{\lambda \lambda'}_{ij,kl}= \int d^3 r  \int d^3 r' \;
       \xi_{i,\lambda}^*(\vec r) \xi_{j,\lambda'}^*(\vec r^{\,\prime}) \;\\
       \times V(\vec r -\vec r^{\,\prime}) \;
       \xi_{k,\lambda'}(\vec r^{\,\prime})\xi_{l,\lambda}(\vec r)
\end{multline}
with the band index $\lambda = e,h$, the single-particle wave function
$\langle \vec r|i, \lambda \rangle = \xi_{i, \lambda}(\vec r)$, and the
Coulomb potential $V(\vec r) = e^2/4 \pi \epsilon_0 \epsilon_r r$,
where $\epsilon_r$ is the background dielectric constant.

It has been shown that in the case of flat, cylindrically symmetric
QDs the single-particle bound-state wave functions in the plane of
larger extension (perpendicular to the growth direction) are well
approximated by those of a two-dimensional harmonic oscillator
\cite{Wojs:96a}. Due to the rotational symmetry around the dot axis the
corresponding angular momentum is a good quantum number. For the
strong confinement in the growth direction we use an infinite
potential well to model the corresponding finite extension of the
wave-function. Only the energetically lowest state due to
confinement in the growth direction will be considered.

A particular situation which is often employed in the literature and
will be adopted in the following is the so-called symmetric case, in
which one assumes identical envelopes for the valence- and
conduction-band electrons, i.e. $\xi_{i,e}=\xi_{i,h}^*=\xi_{i}$. This
assumption holds exactly for QDs modelled by a box-like potential and
is found to be a good approximation for oscillator potentials. Note
that for the same state index $i$ the electrons and holes have
opposite angular momenta.  With the assumption of identical envelopes
all Coulomb matrix elements are related to the electron-electron ones
\begin{equation}
\label{eq:symm}
 V_{ij,kl}^{hh}= V_{kl,ij}^{ee} \quad
 \mathrm{and} \quad V_{ij,kl}^{he}= V_{lj,ki}^{ee} \quad ,
\end{equation}
and the superscript $ee$ of the latter will be skipped in what follows.

The Hamiltonian is fully determined by the effective masses $m_e~=0.065 \, m_0$
and $m_h=0.17 \,  m_0$ for the electrons and holes, respectively, the dielectric
constant $\epsilon_r=13.69$ as well as the oscillator length  $l_{osc}~=5.4 $ nm
(describing the in-plane confinement) and the QD height $L=4$~nm.
From these parameters we obtain the single-particle
energies with constant spacing $\hbar \omega_e={\hbar^2}/{m_e
l_{osc}^2}=40.20$ meV and $\hbar\omega_h=15.37$~meV as well as the energy
scale of the Coulomb matrix elements
$E_c={e^2}/{4\pi\epsilon_0\epsilon_r l_{osc}}= 19.48$~meV. In the
following we will restrict ourselves to a QD containing only $s$- and
$p$-shells for both electrons and holes. In this case the orbital of
the single-particle state $|i \rangle$ can be uniquely identified by
its angular-momentum, $m=0$ for the $s$-shell and $m=\pm 1$ for the
two $p$-states.

The six-fold integral in Eq.~(\ref{eq:Vijkl}) can be analytically reduced to a
one-dimensional integral, which is determined numerically. Due to the
cylindrical symmetry of the problem we have angular momentum conservation:
$V_{ij,kl}\propto \delta_{m_i+m_j,m_k+m_l}$. All non-zero Coulomb matrix
elements are listed in Table~\ref{VijklList}.

\begin{table}[!htb]
\begin{center}
\begin{tabular}{c|c}
    $(i,j,k,l)$ & ${V}_{ij,kl}/E_c$                                   \\ \hline
   (0, 0, 0, 0)                                              & 1.1197 \\
   (0, 1, 1, 0),  (1, 0, 0, 1)                               & 0.8690 \\
   (0, -1, -1, 0), (-1, 0, 0, -1)                            & 0.8690 \\
   (1, 1, 1, 1)                                              & 0.7935 \\
   (1, -1, -1, 1),(-1, 1, 1, -1)                             & 0.7935 \\
   (-1, -1, -1, -1)                                          & 0.7935 \\ \hline
   (0, 0, 1, -1), (0, 0, -1, 1),(1, -1, 0, 0), (-1, 1, 0, 0) & 0.2507 \\ \hline
   (0, 1, 0, 1), (1, 0, 1, 0)                                & 0.2507 \\
   (0, -1, 0, -1),(-1, 0, -1, 0)                             & 0.2507 \\
   (1, -1, 1, -1),(-1, 1, -1, 1)                             & 0.1753

\end{tabular}
\end{center}

\caption{ \label{VijklList} Non-zero Coulomb matrix elements in units of $E_c$. The
indices refer only to the $z$-component of the angular momentum $m=0$
for the $s$-shell and $m=\pm 1$ for the $p$-shell. The horizontal lines
divide the matrix elements into three groups: direct- (top), exchange-
(bottom) and other terms.}
\end{table}

\section{Exact Diagonalization}
\label{fulldiag}

In a semiconductor QD the finite height confinement potential leads to a
finite number of localized states as well as to energetically higher
delocalized states. When the influence of the delocalized states on the
discrete QD spectrum is neglected, the eigenvalue problem has a finite
(albeit large) dimension and can be solved without further approximations.
In this way the Coulomb interaction between the different possible
configurations of carriers in the available bound states is fully taken
into account.

As the Hamiltonian conserves the total number of electrons $N_e$ and
holes $N_h$, the Hamiltonian matrix falls into subblocks with basis
states corresponding to uncorrelated many-particle states of the form
\begin{equation} \label{eq:BasisEH}
| \phi \rangle=\prod_{\substack{i \\ \sum_i n_i^e=N_e}}
       (e^{\dag}_i)^{n^{e}_i}
               \prod_{\substack{j \\ \sum_j n_j^h=N_h}}
       (h^{\dag}_j)^{n^h_j} \,\, |0\rangle \quad .
\end{equation}
Moreover, the Hamiltonian (\ref{eq:ehHamiltonian}) commutes with the
total (electron plus hole) angular momentum $l_z=l_z^e+l_z^h$ , with
the total spin of electrons, ${\bf S}^2_e,S^z_e$, as well as with the
total spin of holes, ${\bf S}^2_h,S^z_h$. This rather rich symmetry
can be used for separating even smaller Hamiltonian subblocks and for
predicting degeneracies. A list of good quantum numbers includes
therefore $N_e,N_h,l_z,{\bf S}^2_e,{\bf S}^2_h,S^z_e,S^z_h$ and the
eigenstates are degenerate with respect to $S^z_e,S^z_h$. By numerical
diagonalization of these subblocks one gets the Coulomb-correlated
states and the corresponding eigenvalues, classified according to the
above-mentioned symmetries.

The coupling of the QD with the optical field is described by the
interband dipole operator
\begin{equation}
\label{eq:P}
  P = \sum_{\sigma}P_{\sigma} = d \sum_{i,\sigma} h_{i,\sigma}e_{i,-\sigma}
\end{equation}
and its hermitian conjugate, where $d$ is the interband dipole matrix
element.  The operator $P$ describes the recombination of
"mirror-symmetric" pairs, i.e., having opposite $z$-components of the
angular momentum and spin.  The diagonality in the orbital index $i$
is a consequence of the identical envelopes of the electron and hole
states. The matrix elements of $P$ define the QD emission as given by
Fermi's golden rule,
\begin{equation}
\label{eq:Fermi}
I(\omega)={2\pi \over \hbar} \sum_f |\langle \phi_f| P |\phi_i \rangle |^2
\,\, \delta(E_i-E_f-\hbar\omega) \quad ,
\end{equation}
where the correlated initial ($i$) and final ($f$) states and their energies are
calculated as described above. The final state has, of course, one
electron-hole pair less than the initial one. The hermitian conjugate
operator $P^{\dag}$ creates mirror-symmetric pairs and appears in a
similar formula for the light absorption.

Next, one has to specify the initial states for which the optical
spectra will be calculated. In the following we restrict ourselves to
situations where optical excitation leads to the same number of
electrons and holes within the QD, $N_e = N_h = N^X$ where $N^X$
stands for the number of electron-hole pairs (in the following loosely
called excitons). It is further assumed that energy relaxation of
carriers within the QD is considerably faster than carrier
recombination such that the initial state for the recombination
process with given $N^X$ is the corresponding multi-exciton state with
the lowest energy ($N^X$ exciton ground state). Moreover, changes of
the carrier spin during relaxation are neglected \cite{Kamada:99}.
Since optical excitation only creates multi-exciton states where
electrons and holes have opposite $S_z$ values, only states with
vanishing total $z$-component of the spin are considered as initial
states in Eq.~(\ref{eq:Fermi}).

The analysis of Coulomb-correlated multi-exciton   states shows that for
even $N^X$, the symmetry of the ground state is singlet-singlet (ss),
i.e., $S_e=S_h=0$. These states are nondegenerate. On the other hand,
for odd $N^X$ one has doublet-doublet (dd) ground states with
$S_e=S_h=1/2$ which are four times degenerate. For the above
discussed choice of the initial states for the recombination process
only two states contribute to the emission formula with a weighting
factor of $1/2$ each. The other two states are also dipole-allowed, but are
eliminated from the emission formula because their total $S_z$ is
nonzero.  (By the same arguments one may be concerned about
eliminating the states not having total angular momentum zero but, as
expected, no such ground states occur.)  The final state can be any
(ground or excited) state of the system with one exciton removed.

\begin{figure}[!htb]
\begin{center}
\includegraphics[width=\columnwidth]{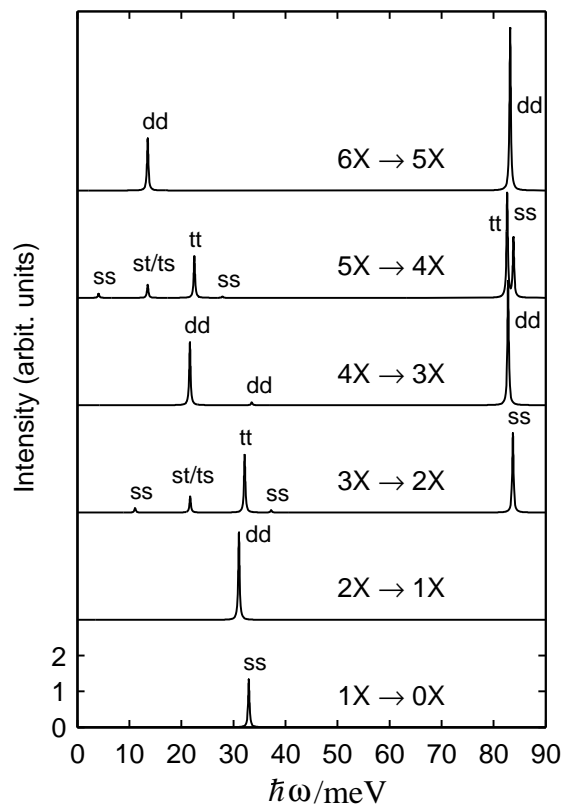}
\caption{
Ground state emission spectra for a quantum dot with different number of excitons. The labels indicate the total
spin of the final state for electrons and holes, respectively: s-singlet, d-doublet, t-triplet. The total angular
momentum $z$-component is always zero.  All energies are measured relative to the band-gap. }
\label{Fig1}
\end{center}
\end{figure}

The results of an evaluation of Eq.~(\ref{eq:Fermi}) based on
Cou\-lomb-correlated multi-exciton states are shown in Fig.~\ref{Fig1}
for an initial filling of one up to six electron-hole pairs. For
better visibility the $\delta$-functions are represented by narrow
Lorentzians centered at the energy $\hbar \omega = E_i-E_f$ (measured
relative to the band gap) and having the area equal to the
oscillator strength $\,|\langle\phi_f|P|\phi_i\rangle|^2 \,$.

Several features are immediately obvious. First, there is a clear
spectral separation between the higher energy $p$-lines for
$\hbar\omega>$ 80 meV (obtained by removing one exciton from the
$p$-shell) and the lower energy $s$-lines (for $\hbar\omega<$~35~meV).
Second, the position of the $s$-lines are arranged approximately
in a descending ladder as the number of excitons increases, while the
energy of the $p$-lines show a remarkable stability. The last fact was
attributed to a 'hidden symmetry' property, which will be discussed
below. The dipole operator $P$ has no simple commutation relation with
the spin operators ${\bf S}^2_e,{\bf S}^2_h$ and therefore the spin
symmetry of the final state is not determined by that of the initial
state \cite{footnote1}. This is why one encounters as final states all
possible spin symmetries (ss, dd, tt, st, ts), as indicated in the
figure.

By restricting the present discussion to $s$- and $p$-shells only, the
Hamiltonian of Eq.~(\ref{eq:ehHamiltonian}) is identical (up to nonessential
differences in the parameters) to the case analyzed in
\cite{Hawrylak:99}. Nevertheless, the full diagonalization procedure used here
leads to different relative line intensities and, more importantly, to the
appearance of new emission lines (the st- and ts-lines are missing in
\cite{Hawrylak:99}).

\section{Approximate Hamiltonians}
The interaction between different state configurations, as given by the exact
diagonalization procedure, shows a quite high degree of complexity and
therefore the results are not immediately intuitive. Even though
we have considered electron-hole pairs that are optically created in
mirror-symmetric states, the Coulomb interaction mixes them strongly with
configurations in which the electrons and holes are {\em not} arranged
symmetrically. For instance, promoting two holes from the $s$-shell to the
$p$-shell is energetically less costly than promoting one electron and one
hole. The second case is more symmetric, but the first produces a state which
is energetically closer and therefore participating stronger in the exact
interacting eigenstate.
This may explain the disagreement with the line intesities found in
\cite{Hawrylak:99}, where only symmetric states are considered.

On the other hand, the relatively regular structure of the emission lines
seems to indicate that an intuitive picture should be possible.
This is achieved by turning to approximate, simpler Hamiltonians which allow
analytic solutions and at the same time retain the essential features of the
full problem.

One such simplified Hamiltonian can be obtained as follows.  By
examining Table \ref{VijklList} one sees that the largest Coulomb
matrix elements are the direct ones, $V_{ij,ji}=D_{ij}=D_{ji}$. Their
contribution to $H_{Coul}$ can be expressed solely in terms of the
occupation number operators, $\hat
n_{i\sigma}^e=e_{i\sigma}^{\dag}e_{i\sigma}$ and $\hat
n_{i\sigma}^h=h_{i\sigma}^{\dag}h_{i\sigma}$, and therefore is diagonal
in the noninteracting basis, Eq.~(\ref{eq:BasisEH}).  The same is true
for the exchange matrix elements $V_{ij,ij}=X_{ij}=X_{ji}$ with $i
\neq j$ provided one includes their contribution only in the e-e and
h-h interaction terms involving the same spin ($\sigma=\sigma'$). In
this way one obtains the diagonal Hamiltonian
\begin{eqnarray}
\label{eq:diag}
H_d &=& \sum_i \left(\varepsilon^e_i-{1 \over 2}D_{ii}\right) \hat n^e_i +
      \sum_i \left(\varepsilon^h_i-{1 \over 2}D_{ii}\right)\hat n^h_i
      \nonumber \\
    &+& {1 \over 2}\sum_{i,j}D_{ij}\left(\hat n^e_i-\hat n^h_i\right)
                                 \left(\hat n^e_j-\hat n^h_j\right)
     \nonumber\\
     &-& {1 \over 2} \sideset{}{'} \sum_{i,j,\sigma}
     X_{ij}\left(\hat n^e_{i,\sigma}\hat n^e_{j,\sigma}+
                  \hat n^h_{i,\sigma}\hat n^h_{j,\sigma}
      \right)  \quad .
\end{eqnarray}
The prime in the last summation indicates that $i=j$ terms have to be
omitted and $\hat n_i^{e,h}=\hat n_{i \uparrow}^{e,h}+\hat n_{i
\downarrow}^{e,h}$. Of course, for this Hamiltonian there is no
configuration interaction. The non-correlated states are eigenstates
and the eigenvalues are derived by the above formula by inserting the
corresponding occupation numbers.

\begin{figure}[htb]
\begin{center}
 \includegraphics[width=\columnwidth]{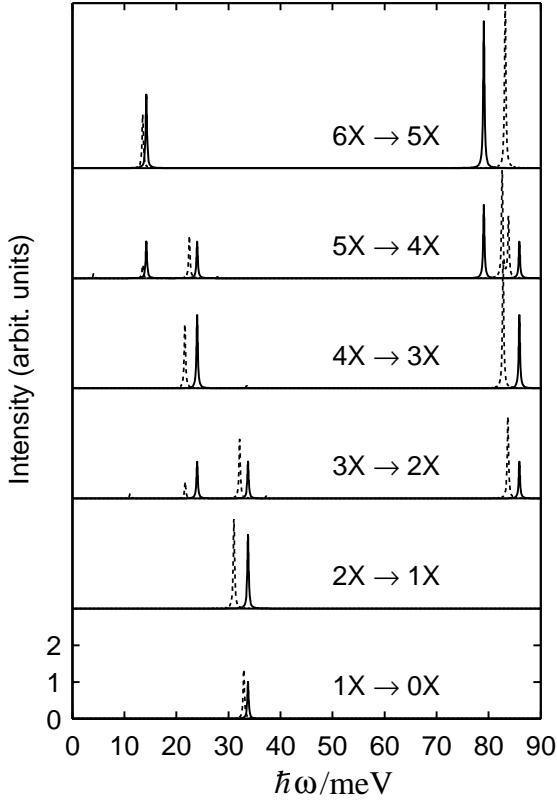}
 \caption{ 
 Comparison of results from the diagonal Hamiltonian, Eq.~(\ref{eq:symdiag}), (solid line) and
 the  full diagonalization (dashed line) for quantum dot emission spectra with
 increasing number of excitons.  }
 \label{Fig2}
\end{center}
\end{figure}

If the states are symmetrically populated,  $n^e_{i \sigma}=n^h_{i,
 -\sigma}=n^X_{i \sigma}$, their energy is further simplified:
\begin{equation}
\label{eq:symdiag}
E = \sum_i \left(\varepsilon^e_i+\varepsilon^h_i-D_{ii}\right)n^X_i
  - 2 \sum_{i<j \atop \sigma}
  X_{ij}\,n^X_{i,\sigma}n^X_{j,\sigma} \quad .
\end{equation}
In this model, the exciton energy is
$E^X_i=\varepsilon^e_i+\varepsilon^h_i-D_{ii}$ where the binding
energy results from the direct electron-hole Coulomb attraction.  When
the electron and hole envelope functions are identical, the excitons
are not only globally but also locally neutral and the direct
electrostatic interaction between different excitons vanishes. In the
approximate Hamiltonian~(\ref {eq:diag}) the only interaction between excitons
comes from the exchange terms and takes place between excitons with the same
spin structure \cite{footnote2}. For instance, the biexcitonic binding
energy is zero in this approximation (the full model gives a small
binding energy of about 2 meV).  A comparison of the diagonal model
with the full result is given in Fig.~\ref{Fig2}. The approximate
spectrum indeed shows a ladder-like structure of the $s$-lines, with the
corresponding spacing in good agreement with the full calculation and
given by the exchange interaction between carriers in the $s$- and
$p$-shell, $2X_{sp}$.  According to Eq.~(\ref{eq:symdiag}) the energy
change for the removal of one $s$-exciton with a given spin structure is
proportional to the number of $p$-excitons having the same spin
structure. Adding equal contributions from electrons and holes the
coefficient is $2X_{sp}$.  Also, in the case of an odd number of
excitons one has two $s$-lines, depending on whether the spins in the
removed $s$-exciton agree or not with the spin orientation in the
majority of the $p$-excitons. For an even number of excitons this
splitting does not occur. These main spectral features have been
described previously \cite{Dekel:98, Dekel:00a}, here
we show which terms of the full Hamiltonian are responsible for them and
that these terms can be included in an exactly solvable approximate
Hamiltonian $H_d$.

Even though the diagonal Hamiltonian gives an intuitive picture of the
main features in terms of uncorrelated states,
the following shortcomings of this model need to be
discussed. The ground state of the 4-exciton problem is inaccurately
given by the diagonal model as a triplet-triplet state, as predicted
by Hund's rule. It is known that Hund's rule does not always apply in
QDs and this is such a case. The true ground state with an energy
slightly below this triplet-triplet state has a singlet-singlet
symmetry, as mentioned in the previous section. It was this
singlet-singlet state which was used as initial state in the emission
spectrum obtained from the diagonal Hamiltonian, solid line in  Fig.~\ref{Fig2},
in order to make the comparison meaningful.  Also the diagonal Hamiltonian 
entails a ladder-like structure for the $p$-lines too, with a spacing of $2X_{pp}$,
which is not confirmed by the full calculation. As discussed below, this stems
from neglecting important pieces from the interaction.

Regarding the $p$-shell emission, one can refine the approximate
Hamiltonian in the following way.  An examination of the ground states
for a situation with three or more excitons, calculated with the full
diagonalization procedure of Section~\ref{fulldiag}, shows that it is safe to
assume that in such states the $s$-shell is completely filled. The
states with full $s$-shell configurations appear with a weighting factor
of at least 0.95. Therefore, as far as the ground states and the lower
excited states are concerned, it is possible to construct an
approximate Hamiltonian describing the fully interacting $p$-states
following 'adiabatically' an external field provided by the 'frozen'
$s$-carriers.  Practically this is obtained from
Eq.~(\ref{eq:ehHamiltonian}) along the same lines as before, but this
time one enforces diagonality {\em only with respect to the
$s$-occupation numbers}. In other words, one discards only those terms
which contain $s$-state creation or annihilation operators and cannot be
expressed in terms of $s$-state occupation numbers. In this way, one
obtains fully correlated $p$-states at given $s$-orbital fillings.

In the resulting 'adiabatic' Hamiltonian $H_{ad}=H_{ad}^{(s)}+
H_{ad}^{(p)}+ H_{ad}^{(sp)}$ we have separated the terms describing
the $s$- and $p$-shell as well as the $s$-$p$-interaction.  The $p$-shell part
retains the form of Eq.~(\ref{eq:ehHamiltonian}) with the summation
restricted to the $p$-orbitals. Therefore, in the following equations,
the indices $i,j,k,l$ label only $p$-states while for the $s$-states the
explicit subscript $s$ is taken.  Using the symmetry relations of the
Coulomb matrix elements, Eq.~(\ref{eq:symm}), and bringing close the
operators with the same spin, $H_{ad}^{(p)}$ can be rewritten as
\begin{eqnarray}
\label{eq:adp}
 H_{ad}^{(p)}&=&
          \sum_{i\sigma}\varepsilon^e_i \; \hat n^e_{i\sigma}
                 - {1 \over 2}\sum_{ijk\sigma}
                 V_{ij,kj} \; e^{\dag}_{i\sigma}e_{k\sigma}  \nonumber \\
     &+&  \sum_{i\sigma}\varepsilon^e_i \; \hat n^h_{i\sigma}
                 - {1 \over 2}\sum_{ijk\sigma}
                 V_{ij,kj} \; h^{\dag}_{i\sigma}h_{k\sigma}  \nonumber \\
     &+& {1 \over 2}\sum_{ijkl \atop \sigma \sigma'} V_{ij,kl}
        \left(e^{\dag}_{i,\sigma}e_{l,\sigma}-
                      h^{\dag}_{l,\sigma}h_{i,\sigma}\right) \nonumber\\
     &~& \hphantom{{}\sum_{i \atop \sigma } V_{ijkl}}
         \times\left(e^{\dag}_{j,\sigma'}e_{k,\sigma'}-
                      h^{\dag}_{k,\sigma'}h_{j,\sigma'}\right)
\end{eqnarray}
where the new one-particle Coulomb terms result from the reordering of
operators. The $s$-shell part and the $s$-$p$-interaction are similar to the
diagonal Hamiltonian, Eq.~(\ref{eq:diag}),
\begin{multline}
\label{eq:ads}
H_{ad}^{(s)}
      = \left( \varepsilon^e_s - {1 \over 2}D_{ss}\right)\hat n^e_s
      +  \left( \varepsilon^h_s - {1 \over 2}D_{ss}\right)\hat n^h_s  \\
      + {1 \over 2}D_{ss}\left(\hat n^e_s-\hat n^h_s \right)^2 \quad ,
\end{multline}
\begin{multline}
\label{eq:adsp}
H_{ad}^{(sp)}
      = D_{sp}\left(\hat n^e_s - \hat n^h_s \right)
              \left(\hat n^e_p - \hat n^h_p \right) \\
      - X_{sp} \sum_ {\sigma}\left(\hat n^e_{s,\sigma} \hat n^e_{p,\sigma}+
                                   \hat n^h_{s,\sigma} \hat n^h_{p,\sigma}
      \right) \quad ,
\end{multline}
where $\hat n^{e,h}_{s,p}$ stands for the total population of
electrons or holes on the $s$- or $p$-shell, respectively.  It is
important to note that, with the Coulomb matrix elements listed in
Table \ref{VijklList}, in the one-particle Coulomb terms of
Eq.~(\ref{eq:adp}) only the contributions with $i=k$ appear. They
are responsible for a renormalization of the one-particle energies
$\varepsilon_i^{e,h}$. Moreover, these renormalized energies do not
depend on the index $i$, so that $H_{ad}^{(p)}$ describes a 'single
degenerate shell' \cite{Wojs:96,Hawrylak:96,Hawrylak:99},
and can be rewritten as
\begin{align}
\label{eq:adp1}
 H_{ad}^{(p)}&=  \left( \varepsilon^e_p - {1 \over 2}D_{pp}- {1 \over
    2}X_{pp}\right)\hat n^e_p    \nonumber \\
      &+  \left( \varepsilon^h_p - {1 \over 2}D_{pp}- {1 \over
    2}X_{pp}\right)\hat n^h_p     \nonumber \\
     &+ {1 \over 2}\sum_{ijkl \atop \sigma \sigma'} V_{ij,kl}
        \left(e^{\dag}_{i,\sigma}e_{l,\sigma}-
                      h^{\dag}_{l,\sigma}h_{i,\sigma}\right) \nonumber  \\
     & \hphantom{{1 \over 2}\sum_{ijkl \atop \sigma \sigma'} V_{ij,kl}}
        \times\left(e^{\dag}_{j,\sigma'}e_{k,\sigma'}-
                      h^{\dag}_{k,\sigma'}h_{j,\sigma'}\right) \quad .
\end{align}

With the help of the adiabatic Hamiltonian $H_{ad}$ the full emission spectrum
can be explained as follows. For the $s$-lines the arguments showing the
formation of a ladder with the spacing of $2X_{sp}$ are as in the case of the
diagonal Hamiltonian $H_d$. The energetic position of the $p$-shell emission can
be deduced from the commutation relation of $H_{ad}$ with the $p$-shell
dipole-transition operator $P_p$, defined as in Eq.~(\ref{eq:P}) but with the
sum restricted only to $p$-states. It is readily verified, that
\begin{equation}
\label{eq:hidden}
\left[e^{\dag}_{i,\sigma}e_{j,\sigma}-
 h^{\dag}_{j,\sigma}h_{i,\sigma},P_p \right] = 0
\end{equation}
and therefore $P_p$ commutes with the last two lines of
Eq.~(\ref{eq:adp1}). This is the core of the 'hidden symmetry'
property \cite{Wojs:96,Hawrylak:96} showing that the interaction part
plays no
role in this argument. The commutation of $P_p$ with the occupation
number operators of the shells is rather obvious, because the
application of $P_p$ corresponds to a population reduction by one
electron-hole pair in the p-states and no change in the
s-states. Correspondingly, one can verify $\left[\hat n^{e,h}_p,
P_p\right]= -P_p$ and $\left[\hat n^{e,h}_s, P_p\right]= 0$.  Such
simple relations arise only when commuting $P_p$ with the total number
operator of the $p$-shell, not with individual number operators. This is
why it is important to have degenerate shells.

Assuming, that the s-states are fully occupied  
($n_{s,\sigma}^{e,h}~=~1$) one obtains from these results
\begin{equation}
\left[ H_{ad},P_p \right] = -\left( \varepsilon^e_p + \varepsilon^h_p- D_{pp}-
  X_{pp} - 2 X_{sp}\right) P_p \quad .
\end{equation}
This shows that the removal of one $p$-exciton is accompanied by an energy
decrease which does not depend on the number of excitons. The value of this
energy,
\begin{align}
\label{eq:noladder}
\mathcal{E}_p^X &= \varepsilon^e_p + \varepsilon^h_p- D_{pp}-X_{pp} - 2 X_{sp} 
\nonumber \\
&= E_p^X- X_{pp} - 2 X_{sp} \quad ,
\end{align}
is in excellent agreement with the exact diagonalization result. Indeed, using
the values in Table~\ref{VijklList} one obtains
$\mathcal{E}_p^X=82.5\, \mathrm{meV}$. The emission spectrum associated with the
adiabatic Hamiltonian consists therefore of $p$-lines having all this
common value, while the $s$-line ladder remains the same as given by the diagonal Hamiltonian.

The 'hidden symmetry' argument, as discussed by Wojs and
Hawrylak, \cite{Wojs:96, Hawrylak:96, Hawrylak:99} is proven on
the assumptions that (i) one has identical envelopes in the two bands and (ii)
the one-particle levels form 'a single degenerate shell'.
In these hypotheses one gets a constant energy value at the removal of each
exciton, and this value depends only on the parameters of this shell.

Nevertheless, the Hamiltonian under discussion here, as generally used
in the literature, is rather describing several interacting shells, so that
is not obvious why (if at all) the argument holds. In the present case
the answer is contained in the adiabatic Hamiltonian. In it the $p$-shell is the
'single degenerate shell' because (i) the $s$-shell is `frozen' and higher shell
are absent in the considered situation and (ii) the field created by the
$s$-shell carriers does not remove the degeneracy of the $p$-shell. In such
conditions the adiabatic Hamiltonian obeys exactly the 'hidden symmetry'
commutation relations. In this picture the energy for the removal of a
$p$-exciton, Eq.~(\ref{eq:noladder}), contains also terms coming from the
$s$-$p$-interaction, and it is this value that is in agreement with the full
diagonalization.

This discussion also shows the validity range of the 'hidden symmetry'. For
instance, the presence of higher shells (but also depending on the
actual model parameters) may spoil the argument. This seems to be the case
described in \cite{Dekel:00a}, where the $p$-lines are not independent on the
exciton number, but are arranged in a descending ladder too. Also, the field of
the 'frozen' states may remove some of the degeneracy of the shell in
question. For example if the outer shell is a $d$-shell, the states with zero
angular momentum and those with angular momentum $\pm 2$ will experience
the field created by the $s$-shell carriers differently .

One may argue that by approximating the full Hamiltonian,
Eq.~(\ref{eq:ehHamiltonian}), with the adiabatic one,
Eqs.~(\ref{eq:adp})-(\ref{eq:adsp}), the interaction inside the
$p$-shell is still too complicated to allow analytic diagonalization,
i.e. the 'hidden symmetry' property is a simple relation between otherwise
complex, strongly correlated states.  Nevertheless, the Fock
subspace generated by the $p$-orbitals is significantly smaller and
this is in itself a numerical simplification.
Symmetry arguments also can be used to reduce the blocks to be
diagonalized and at least the ground states can be obtained
analytically. A procedure for obtaining analytical eigenstates is
the repeated applications of the raising operator $P^\dag_p$ on the
'vacuum' (full $s$-shell, empty $p$-shell) state
\cite{Wojs:96,Hawrylak:96,Hawrylak:99}.

In summary, for semiconductor QDs with finite height two-dimensional
harmonic confinement potential the multi-exciton emission spectra are
discussed on the basis of a full diagonalization of the Hamiltonian
including Coulomb interaction for the localized states.  The
characteristic features are $s$-shell lines arranged approximately in a
descending ladder with increasing exciton number as well as nearly
constant energetic position of the $p$-shell lines, provided that these
are the only confined shells.  Based on the relative importance of the
various Coulomb matrix elements, a simplified Hamiltonian has been
constructed which is diagonal in the single-particle states. It
explains the $s$-shell emission properties, that appear as soon as the
$p$-shell population starts to contribute, in terms of the $s$-$p$-exchange
interaction $X_{sp}$. The results are two main $s$-shell lines for odd
number of excitons separated by $2X_{sp}$, while the spectrum is
dominated by a single line for an even exciton number.  The almost
constant energetic position of the $p$-shell emission ('hidden
symmetry') is discussed in terms of Coulomb correlated $p$-shell
carriers in the presence of a completely filled $s$-shell. It is also
shown that the arguments for the 'hidden symmetry' break down as soon
as higher confined shells contribute. Depending on the coupling
strength of higher shells to the $p$-shell the energetic stability of
the $p$-lines is expected to disappear gradually.

The proposed simplified Hamiltonians give a more intuitive picture of
the rich properties of the emission spectra from the
Coulomb-correlated QD carriers. They also might be an alternative to a
demanding full diagonalization scheme when Coulomb interaction in the
presence of three and more electron-hole pairs is important. Finally,
only a diagonal Hamiltonian allows to discuss physical processes in
terms of single-particle states. Our comparison of results from full
and diagonal Hamiltonian shows to what extent this is justified.

\section*{Acknowledgments}

We thank P. Hawrylak for valuable discussions. This work has been
financially supported through the Deutsche   Forschungsgemeinschaft. A
grant for CPU time at the Forschungszentrum J\"{u}lich is gratefully
acknowledged.


\begin{thebibliography}{99}
\bibitem{footnote2} Since we only consider electron-hole pairs with
opposite $z$-components of the spin, the exciton spin is always
zero. Exchange interaction between two excitons is possible when the
$z$-components of the spin for the two electrons agree, which is
then also true for the holes.
\end{thebibliography}
\end{document}